\documentstyle[12pt]{article}
\begin{document}
\begin{titlepage}
\vspace*{-2cm}
\centerline{\bf Near-Threshold Production of $\eta$-Mesons}
\vspace*{1cm}
\centerline{Colin Wilkin}
\vspace*{1cm}
\centerline{University College London, London WC1E 6BT, U.K.}
\begin{abstract}
It is shown that the striking energy variation in the $p\, d \rightarrow
\mbox{$^{\, 3}{\rm He}$}\, \eta$  cross section near threshold is probably due
to a final state interaction associated with a large (complex)
$\eta -\mbox{$^{3}{\rm He}$}$ scattering length. The consequences of this
hypothesis are studied for the production of the meson in the
$\eta -\mbox{$^{4}{\rm He}$}$ and $\eta -\mbox{$^{7}{\rm Be}$}$ channels.
\end{abstract}
\vspace*{2cm}
PACS number(s): 25.40.Ve, 13.75.-n, 25.10.+s
\end{titlepage}

The cross sections for the production of $\eta$ and $\pi^0$ mesons {\em via}
the
$p\, d \rightarrow \mbox{$^{3}{\rm He}$}\, \eta$  and
$p\, d \rightarrow \mbox{$^{3}{\rm He}$}\, \pi^{0}$  reactions behave very
differently close to their respective thresholds. Taking out the kinematic
factor of the ratio of the outgoing to the incident centre-of-mass momenta, we
can define an average squared amplitude through
\begin{equation}
|f_{\eta (\pi)}|^{2} = \frac{p_{p}}{p_{\eta (\pi)}}\,\left(\frac{d\sigma}
{d\Omega}\right)_{\rm cm}\: .
\end{equation}
For $\eta$-production, $|f_{\eta}|^{2}$ decreases by over a factor of three
between threshold and an $\eta$ centre-of-mass momentum of
$p_{\eta}=70\:{\rm MeV/c}$, while remaining essentially independent of
production angle \cite{Berger,Garcon}. This corresponds to a change in beam
energy of less than 10~MeV. In contrast, $|f_{\pi}|^{2}$ shows a
strong angular variation, with the ratio of the forward to backward cross
section already attaining a factor of three by $p_{\pi}=20\:{\rm MeV/c}$.
However the angular average has a much weaker energy dependence than
for the $\eta$ case \cite{Pickar,Kerboul}.

This striking difference in the angular distribution is almost certainly due to
the basic meson-nucleon interaction. The \mbox{$\pi N$} interaction is
governed by a \underline{P-wave} resonance, the $\Delta$(1232), with only a
very weak S-wave. The strong angular dependence seen in the
$p\, d \rightarrow \mbox{$^{3}{\rm He}$}\, \pi^{0}$ cross section is then
a result of an interference of the large P-wave with an S-wave which is only
significant within a few MeV of threshold \cite{GW1}. On the other hand the
most prominent feature of the low energy \mbox{$\eta\, N$} interaction is an
\underline{S-wave} resonance, the N$^{*}$(1535), so that the low energy
$\pi^{-}\, p \rightarrow \eta\, n$  reaction shows only a comparatively weak
angular dependence \cite{Binnie}. It is our contention that this strong S-wave
interaction is also responsible for the rapid energy variation of the
near-threshold $p\, d \rightarrow \mbox{$^{3}{\rm He}$}\, \eta$  cross section.

Whereas the $p\, d \rightarrow \mbox{$^{3}{\rm He}$}\, \pi^{0}$ cross section
and deuteron tensor analysing power are both well described at low energies by
a model involving a spectator nucleon \cite{GW1}, such an approach fails for
the $p\, d \rightarrow \mbox{$^{3}{\rm He}$}\, \eta$  reaction and
three-nucleon
mechanisms have been suggested as the origin \cite{LL,GW2}. Though estimates of
the effects of such terms within a semi-phenomenological model do reproduce
some of the features of the data, they predict little energy dependence of
the amplitude within a few MeV of threshold \cite{Kilian,CW}. It should though
be noted that these models are perturbative, treating all interactions only to
lowest order, and this might not be justified for low energy $\eta$-nucleon
scattering where the S-wave is very strong.

A common approximation \cite{Gold}, in the case of a weak transition to a
channel with a strong final-state interaction (FSI), yields an S-wave threshold
enhancement factor of the amplitude f,
\begin{equation}
f = \frac{f_B}{p\, a\,{\rm cot}\delta - ip\, a}\ \ ,
\end{equation}
where the amplitude $f_B$ is slowly varying near threshold, and $\delta$ and
$a$ are the S-wave phase shift and scattering length in the exit channel, for
which the centre-of-mass momentum is $p$. At low energies it is often
sufficient to take
\begin{equation}
f \approx \frac{f_B}{1 - ip\, a}\ \ .
\end{equation}
The approximation leading to this expression corresponds to imposing unitarity
with \underline{constant} K-matrix elements, {\em i.e.} neglecting effective
range effects \cite{Dalitz}. In view of our dearth of knowledge about the low
energy $\eta$-nucleon (nucleus) interaction, it is pointless trying to go
further at present. It should be noted that the effects of the S$_{11}$
resonance are felt in the final state interaction factor rather than in the
$f_B$ term.

Bhalerao and Liu \cite{Liu} analysed the $\pi N$ and $\eta N$ coupled channels
around the $\eta$ threshold within an isobar model and extracted a value for
the $\eta$-nucleon scattering length of
$a\, (\eta N) = (0.27 + i0.22)\,{\rm fm}$.
However a value which is more consistent with our later use of it may be
obtained by applying eq.~(3) directly to $\pi^{-}\, p \rightarrow \eta\, n$
data.

Using detailed balance and the optical theorem, a lower bound on the imaginary
part of $a\, (\eta N)$ is provided by the threshold $\pi^{-}\, p \rightarrow
\eta\, n$  cross section:
\begin{equation}
{\rm Im}[a\, (\eta N)]\ \geq\ \frac{3}{8\pi}\: \frac{p_{\pi}^{2}}{p_{\eta}}\:
\sigma_{\rm tot}(\pi^{-}\, p \rightarrow \eta\, n)\: .
\end{equation}
The data of reference \cite{Binnie} require
${\rm Im}[a\, (\eta N)] \geq (0.28 \pm 0.04)\ {\rm fm}$ which, though a little
larger than the value extracted in \cite{Liu}, is compatible with it. Other
channels, such as $N\pi\pi$, are also open at the $\eta$ threshold and these
must add to the inelasticity \cite{PDG}. We therefore take
${\rm Im}[a\, (\eta N)] = 0.30\ {\rm fm}$, though this might be an
underestimate.

Even after neglecting effective range effects, the
$\pi^{-}\, p \rightarrow \eta\, n$  energy dependence is not sufficient to
determine both the real and imaginary parts of the scattering length but once
the imaginary part is fixed from the transition strength, the fit shown in
fig.~1 leads to
\begin{equation}
a\, (\eta N) = (0.55 \pm 0.20 + i0.30)\ {\rm fm}\: .
\end{equation}
The sign of the real part is ambiguous and we have chosen it to be attractive
to be consistent with the work of Bhalerao and Liu \cite{Liu}. It should be
noted that, though the magnitude is twice the value they found, any P-wave
contributions to the cross section would have the effect of reducing the slope
in energy and hence giving a too low an estimate for $Re[a\, (\eta N)]$.

In impulse approximation the $\eta\,\mbox{$^{3}{\rm He}$}$ scattering length
is essentially three times that of $\eta N$ but there are large corrections to
this simple ansatz due to the strength of the interaction. A more reliable
starting point is to consider the lowest order \mbox{$\eta\,^{3}{\rm He}$}
optical potential for which
\begin{equation}
2m_{\eta\, N}^{R}\: V_{\rm opt}(r) = -4\pi\, A\,\rho (r)\,
a\, (\eta\, N)\: ,
\end{equation}
where $m_{\eta\, N}^R$ is the $\eta$-nucleon reduced mass and A (= 3) the mass
number. Resolving the variable phase equation \cite{Calogero} for this
potential, using a Gaussian nuclear density corresponding to an rms radius of
1.9 fm, leads to a scattering length of
\begin{equation}
a\, (\eta\,\mbox{$^{3}{\rm He}$}) = (-2.31 + i2.57)\ {\rm fm}\: .
\end{equation}
The sign of the real part indicates the possible presence of a `bound' $\eta$
state for a much lighter nucleus than that found by Haider and Liu
\cite{Haider}, but the large imaginary component in the scattering length
limits
its possible significance.

The prediction of the shape of the energy dependence of the
$p\, d \rightarrow \mbox{$^{3}{\rm He}$}\, \eta$  cross section
using the simplified FSI formula of eq.~(3) is shown as the dashed line in
fig.~2, where it is compared with the pioneering SPES4 data \cite{Berger} and
the preliminary SPES2 values \cite{Garcon} which were used to fix the overall
normalisation parameter. The lowest SPES2 point was taken at an average energy
only 200~keV above threshold and is subject to large systematic
uncertainties due to the width of the beam. The energy loss in the target alone
was up to 270~keV \cite{Garcon} and this influences both the value of
$d\sigma /d\Omega$ and $p_{\eta}$ in eq.~(1). If we exclude this doubtful point
then the agreement with experiment, which is stable to modest changes in
${\rm Im}[a\, (\eta N)]$, is impressive. Though numerically a little
fortuitous,
in view of the corrections which might be important to the lowest order optical
potential, nevertheless it indicates that the rapid fall of
the amplitude with energy might indeed be associated with a strong
$\eta\,\mbox{$^{3}{\rm He}$}$ final state interaction.

Once we have the potential of eq.~(6) then we can calculate the phase shift at
all energies, which enables us to use the more general formula of eq.~(2)
rather than the constant scattering length version of eq.~(3). This in fact
makes very little difference, as can be seen from the solid line in fig.~2.

It is not possible to extract directly values of the real ($a_R$) and imaginary
($a_I$) parts of the $\eta\,\mbox{$^{3}{\rm He}$}$ scattering lengths
independently from the present data using eq.~(3). Taking only the highest 6
SPES2 points \cite{Garcon}, a $\chi^2$ minimisation shows that these parameters
are roughly correlated in the form
\begin{equation}
a_{R}^{\, 2} + 0.449 a_{I}^{\, 2} + 4.509 a_{I} = 21.44\: .
\end{equation}
This at least demonstrates that either the real or imaginary part of the
scattering length has to be very large.

A large scattering length, associated with a `bound'
$\eta\,\mbox{$^{3}{\rm He}$}$ system, also seems to be required \cite{Wilkin}
to
explain the cusp-like structure seen for near-threshold production of
states $X$ in the $p\, d \rightarrow \mbox{$^{\, 3}{\rm He}$}\, X$ reaction
for masses close to that of the $\eta$-meson \cite{Plouin}.

In their microscopic model, Laget and Lecolley \cite{LL} include only a
small amount of $\eta$-nucleon rescattering and as a consequence underestimate
severely the energy dependence near threshold. Taking both S and D states in
the nuclear wave function, their graphs are consistent with an $\eta ^3$He
scattering length of modulus 1.6~fm. This leads to only a 25\% decrease in
$|f|^2$ by $p_{\eta}$ = 0.35~fm$^{-1}$ as compared to the factor of three shown
in fig.~2.

The S-wave FSI enhancement factor of eq.~(2) is independent of the entrance
channel though the particular nuclear reaction would influence the amount of
P and higher waves present. It should therefore be applicable also to the
$\pi^{-}\,\mbox{$^{3}{\rm He}$}\rightarrow \eta\,\mbox{$^{3}{\rm H}$}$
reaction. Unfortunately the lowest energy for which this has been measured
\cite{Peng} corresponds to $p_{\eta} = 0.41\ {\rm fm^{-1}}$, which is just off
the scale of fig.~2~!  This may be why Liu \cite{Liu2} did not note any
significant FSI distortion, though it must be stressed that his effective
$\eta$-nucleus potential was also rather weaker. Our analysis indicates that
it could be very interesting to continue the experiment closer to threshold.

The success of our simple interpretation encourages us to look for other
nuclear reactions in which $\eta$-mesons are produced coherently. Data exist
in the case of  $d\, d \rightarrow \mbox{$^{4}{\rm He}$}\, \eta$, but only
away from the threshold region \cite{LG}. Taking an rms radius of 1.63 fm, the
$\eta\,\mbox{$^{4}{\rm He}$}$ potential is stronger but of shorter range than
that for $\eta\,\mbox{$^{3}{\rm He}$}$ and this `binds' further the highly
inelastic $\eta\,\mbox{$^{4}{\rm He}$}$ state. The predicted scattering length
of $(-2.00 + i0.97)\ {\rm fm}$ corresponds to a somewhat less steep energy
dependence than that found for $\eta\,\mbox{$^{3}{\rm He}$}$ production.
Including also effective range effects through eq.~(2), the decrease in $|f|^2$
between $p_{\eta}$ = 0.1 and 0.4 fm$^{-1}$ is expected to be about 2.8 for
$\mbox{$^{3}{\rm He}$}$ but only 1.9 for $\mbox{$^{4}{\rm He}$}$.

The only other case where coherent $\eta$ production on nuclei has been studied
is that of $p\,^{\, 6}{\rm Li}\, \rightarrow \eta\,\mbox{$^{7}{\rm Be}^{*}$}$
\cite{Beppe}, though the energy resolution obtained by detecting the $\eta$
through its 2$\gamma$ decay mode was insufficient to isolate individual states
in the final $^{7}$Be nucleus. Since the optical potential of eq.~(6) predicts
a scattering length of $(-2.92 + i1.21)$~fm and the typical $\eta$
centre-of-mass momentum in this experiment was
$p_{\eta}\sim 0.5\ {\rm fm}^{-1}$, these data lie \underline{outside}
the FSI peak. It might be advantageous to study the cross sections at say
1--2 MeV above the threshold for the excitation of a particular level.

In summary, the very simplified analysis presented here shows that the strong
energy dependence associated with coherent $\eta$ production on nuclei is
consistent with a large (complex) $\eta$-nucleus scattering length. Theoretical
models of near-threshold production which ignore FSI effects must therefore be
treated with caution.

Valuable discussions with L.Castillejo, A.M.Green and S.Wycech were much
appreciated as was the advice from M.Gar\c{c}on and R.Kessler on the results
of ref.~\cite{Garcon}. Continued computational assistance of G.J.Crone was
also valuable.

\newpage\noindent
Figure 1: {The square of the $\pi^{-}\, p \rightarrow \eta\, n$  amplitude
defined by eq.~(1), and extracted from the total cross section data of
ref.~\cite{Binnie}, as a function of the $\eta$ centre-of-mass momentum
$p_{\eta}$. The solid line is a fit using eq.~(3) with the imaginary part of
the $\eta N$ scattering length constrained by unitarity. This leads to the
parameters of eq.~(5). The dashed line is the best fit with
${\rm Im}[a\, (\eta N)] = 0$.}\\[5ex]
Figure 2: {The square of the
$p\, d \rightarrow \mbox{$^{3}{\rm He}$}\, \eta$  amplitude defined by eq.~(1)
and extracted from the cross section data of SPES2 \cite{Garcon} (circles) and
SPES4 \cite{Berger} (crosses) as a function of the $\eta$ centre-of-mass
momentum $p_{\eta}$. The lowest SPES2 point is subject to large systematic
errors due to beam width effects, including energy losses in the target. The
dashed curve is the prediction of eq.~(3) with the scattering length of eq.~(7)
derived from an optical potential. The solid curve is based on eq.~(2)
and includes effective range effects. In both cases the overall normalisation
is a free parameter.}
\end{document}